\documentclass[journal]{IEEEtran}
\usepackage[utf8]{inputenc}
\usepackage[T1]{fontenc}
\usepackage[english]{babel}
\usepackage{graphicx}
\usepackage{booktabs}
\usepackage{multirow}
\usepackage{url}
\usepackage{hyperref}
\usepackage{tabularx}
\usepackage{array}
\usepackage{amsmath}
\usepackage{longtable}
\usepackage{makecell}
\hypersetup{hidelinks}
\usepackage{rotating}
\newcolumntype{L}[1]{>{\raggedright\arraybackslash}p{#1}}
\newcolumntype{C}[1]{>{\centering\arraybackslash}p{#1}}
\newcolumntype{R}[1]{>{\raggedleft\arraybackslash}p{#1}}
\title{Mapping Quantum Threats: An Engineering Inventory of Cryptographic Dependencies}

\author{
  \IEEEauthorblockN{Carlos Benitez\IEEEauthorrefmark{1}}\\
  \IEEEauthorblockA{\IEEEauthorrefmark{1}\textit{Platinum Ciber} --- \texttt{carlos@platinumciber.com}}
}

\begin{document}
\maketitle

\begin{abstract}
The prospective emergence of large-scale quantum computers capable of executing Shor’s algorithm at cryptographically relevant scale would render widely deployed public-key cryptography computationally insecure. Under this threat model, both confidentiality of previously protected data and the authenticity of digital signatures could be compromised across multiple layers of digital infrastructure.

This paper presents a systematic engineering inventory of technologies that depend on quantum-vulnerable asymmetric cryptography. The analysis is structured along two complementary axes—technology domain and operational environment—linking cryptographic primitives to their real-world deployment contexts. The resulting framework provides a structured basis for identifying systemic exposure to quantum-related risks across contemporary digital ecosystems.
\end{abstract}


%--------------------------------------------------------------------------------------------
\section{Introduction}
Modern digital infrastructure relies on asymmetric cryptography for identity verification, secure key establishment, and data integrity. The potential realization of large-scale quantum computing would invalidate the computational hardness assumptions underlying these protocols. Specifically, the susceptibility of integer factorization and discrete logarithm problems to quantum-accelerated algorithms renders widely deployed RSA, Diffie-Hellman (DH), and Elliptic Curve Cryptography (ECC) primitives insecure \cite{shor1994, grover1996}. These vulnerabilities compromise both the retrospective confidentiality of intercepted traffic and the prospective authenticity of digital signatures (see Section~\ref{subsec:quantum-threat-models}).

In response the U.S. National Institute of Standards and Technology (NIST) and the Internet Engineering Task Force (IETF) have standardized initial Post-Quantum Cryptography (PQC) algorithms and hybrid deployment frameworks \cite{fips203, fips204, fips205, ietf-hybrid-tls}. However, while the algorithmic transitions are well-defined, the practical engineering challenge of identifying systemic dependencies remains under-documented at the system-dependency level.

Much of the existing literature remains abstract, focusing on theoretical security reductions or timeline estimations \cite{mosca2018, enisa-quantum}. While recent efforts have introduced protocol-focused threat taxonomies \cite{cryptoeprint:2025/1668}, there remains a lack of granular mapping across heterogeneous infrastructures and operational environments. This paper addresses that gap by presenting a cross-domain inventory of quantum risks, bridging the divide between theoretical cryptographic weaknesses and the realities of deployed system architectures.

The inventory is presented from two complementary viewpoints. The first is a technology-domain perspective, which groups cryptographic components by functional role irrespective of platform. The second is an environment-based perspective, detailing vulnerabilities across major operating environments.

The remainder of this paper is structured as follows. Section~\ref{sec:background} provides the theoretical background on quantum threats and the current status of PQC standardization. Section~\ref{subsec:quantum-threat-models} defines the primary risk vectors—HNDL and signature forgery—and identifies the classical algorithms susceptible to each. Section~\ref{sec:related-work} evaluates existing literature and taxonomies to contextualize the contribution of this inventory. Section~\ref{sec:methodology} then outlines the taxonomy and methodology used to construct the mapping. The core analysis is implemented across the technology-domain and operational-environment views, with findings consolidated in Table~\ref{tab:techdomains} and Table~\ref{tab:op-environments}, respectively. Later, Section~\ref{sec:conclusion} concludes the paper and Section~\ref{sec:future-work} presents a discussion of future research directions.
%------------------------------------------------------------------------
\section{Background}
\label{sec:background}
Classical public-key cryptography derives its security from problems believed to be computationally intractable for conventional computers. The RSA cryptosystem relies on the hardness of integer factorization, while Diffie--Hellman (DH) and its elliptic-curve variant (ECDH) rely on the discrete logarithm problem in finite fields and elliptic-curve groups, respectively. Digital signature schemes such as RSA-PKCS\#1, ECDSA, and EdDSA similarly depend on the infeasibility of computing modular or elliptic-curve discrete logarithms. These assumptions would be invalidated in the presence of large-scale quantum computers.

\subsection{Quantum Algorithms and Their Implications}
Shor’s algorithm demonstrated that integer factorization and discrete logarithms can be solved in polynomial time on a sufficiently powerful quantum computer \cite{shor1994}. This directly breaks RSA, DH, and ECC, thereby undermining nearly all deployed asymmetric cryptography. Grover’s algorithm, though more limited, provides a quadratic speed-up for unstructured search problems \cite{grover1996}. Under standard assumptions, Grover’s algorithm reduces the effective brute-force security margin of symmetric primitives from n bits to approximately n/2 bits (e.g., AES-128 to ~64-bits).

\subsection{Quantum Threat Models}
\label{subsec:quantum-threat-models}
Two distinct but complementary threat models are relevant when analyzing the quantum risk surface:
\begin{itemize}
    \item \textbf{Harvest Now, Decrypt Later (HNDL):} An adversary records encrypted communications now and stores the ciphertexts. Once a cryptographically relevant quantum computer is available, the adversary uses Shor’s algorithm to recover session keys from DH/ECDH or RSA key exchanges, retroactively decrypting the data. This threat model applies directly to a great number of public key protocols.
    \item \textbf{Forgery of Digital Signatures and Certificates:} Breaking RSA, ECDSA, or EdDSA allows an adversary to generate valid signatures without knowledge of the private key, just from the public key. This enables creation of counterfeit digital certificates.
\end{itemize}

\subsection{Post-Quantum Cryptography Standardization}
In response, the U.S. National Institute of Standards and Technology (NIST) initiated the Post-Quantum Cryptography (PQC) project in 2016 \cite{nistpqc}. In 2022, NIST selected the first algorithms for standardization, finalized as Federal Information Processing Standards (FIPS) in 2024: ML-KEM \cite{fips203}, ML-DSA \cite{fips204}, and SLH-DSA \cite{fips205}. Falcon has since been developed into FIPS 206 (FN-DSA), targeted at constrained environments where bandwidth is limited. In parallel, the Internet Engineering Task Force (IETF) has initiated work on \emph{hybrid cryptography}, combining classical algorithms with PQC to enable near-term deployment \cite{ietf-hybrid-tls,ietf-hybrid-ike}.

\subsection{Industrial and Regulatory Context}
Beyond the Internet, critical infrastructure and vertical sectors (e.g., energy, telecommunications, finance, and automotive) are equally exposed. Standards bodies such as ETSI \cite{etsi-qsc}, ENISA \cite{enisa-quantum}, and IEC \cite{iec62351} have initiated quantum-safety analyses. In the automotive domain, secure firmware updates and V2X communications are governed by initiatives such as Uptane \cite{uptane-spec}, IEEE~1609.2 \cite{ieee1609-2}, and ETSI ITS \cite{etsi-its}, all of which depend on digital signatures vulnerable to quantum attacks. Vendors including Microsoft, Apple, Google, and Linux distributions have started deploying hybrid key exchanges in TLS, SSH, and messaging protocols.

%------------------------------------------------------------------------------
\section{Related Work}
\label{sec:related-work}
Previous literature on quantum risk has largely focused on three areas: cryptanalytic timelines, high-level risk assessments, and protocol-specific evaluations. Mosca \cite{mosca2018} established foundational probabilistic timelines for quantum realization, providing a strategic framework for migration urgency. High-level infrastructural risks have been documented by ENISA \cite{enisa-quantum} and ETSI \cite{etsi-qsc}, which identify broad sectors vulnerable to quantum-enabled threats. At the protocol level, Baseri et al. \cite{baseri2024quantumnetwork} provided a technical review of quantum-readiness for standard network protocols, including TLS, IPsec, SSH, and PGP, identifying the specific asymmetric primitives within these stacks that require replacement.

The most closely related work to this paper is Egbuagha and Ikwunna (2025) \cite{cryptoeprint:2025/1668}, which presents a survey of quantum threats and post-quantum strategies organized by cryptographic functionality. Their analysis introduces a domain-based taxonomy spanning key exchange, digital signatures, and secure messaging. However, their work does not explore environmental deployment contexts or provide implementation-level threat mappings for deployed technologies.

In contrast, the present work introduces a dual-axis inventory of quantum risks. It bridges the gap in the literature by mapping vulnerabilities not only by technology domain but also by real-world operational environment. By explicitly cataloging affected implementations and classifying risk vectors (HNDL and signature forgery) this work provides a comprehensive engineering "threat map" that accounts for both architectural and environmental dependencies.

%------------------------------------------------------------------------
\section{Methodology}
\label{sec:methodology}
This study aims to construct a unified inventory of real-world systems and protocols critically exposed to quantum threats. The approach combines documentary analysis of technical specifications with a cross-domain classification of vulnerable technologies.

\subsection{Sources of Evidence}
The inventory is based on publicly available, authoritative sources:

\begin{itemize}
    \item \textbf{Standards bodies:} Specifications and guidance from IETF, NIST, ETSI, ENISA, and IEC.
    \item \textbf{Vendor and platform documentation:} Technical papers and protocol details from major OS vendors, cloud providers, and widely used open-source projects.
    \item \textbf{Academic and industrial research:} Peer-reviewed studies and reports on quantum vulnerabilities in cryptographic protocols, digital identity, software distribution, and critical infrastructure.
\end{itemize}

\subsection{Classification Criteria}
Each entry is evaluated along two dimensions:
\begin{enumerate}
    \item \textbf{Cryptographic primitives:} Identification of asymmetric algorithms in use.
    \item \textbf{Quantum threat vector:} Categorization as either \emph{HNDL}  or \emph{Forgery}, depending on whether confidentiality or authenticity is compromised.
\end{enumerate}

\subsection{Inventory Structure}
The results are organized in two complementary views:
\begin{itemize}
    \item A \textbf{technology-domain view} grouping cryptographic functions regardless of platform.
    \item An \textbf{operational environment view} highlighting how vulnerabilities manifest in specific deployment contexts.
\end{itemize}

\subsection{Scope and Limitations}
This work focuses solely on vulnerabilities introduced by quantum attacks against asymmetric cryptography. Symmetric algorithms and hash functions are assumed quantum-resistant when using appropriate key lengths.

\subsection{Research Questions}
The study is guided by the following questions:
\begin{itemize}
    \item Which systems, platforms, and protocols depend on public-key algorithms vulnerable to quantum attacks?
    \item How do HNDL and forgery threats manifest across different technology domains and environments?
\end{itemize}

%------------------------------------------------------------------------
\section{Inventory of Quantum Risks}
\noindent This section presents the unified inventory of systems affected by quantum threats, following the organizational model introduced in Section~\ref{sec:methodology}. Two complementary views are provided: (i) a technological domain-oriented perspective (Table~\ref{tab:techdomains}), which highlights cross-cutting cryptographic building blocks; and (ii) an operational environment-oriented perspective (Tables~\ref{tab:op-environments}), which illustrates how these risks materialize within concrete ecosystems where cryptography is deployed in practice.

The following subsections expand on each perspective in detail, providing explanatory context for the items listed in the tables.

% * * * * * * * *
\subsection{Technologies by Domain}
This perspective organizes vulnerable systems according to their functional role in the digital ecosystem. By grouping technologies into domains that represent the most widely deployed and security-critical functions, it becomes possible to identify common cryptographic dependencies and attack surfaces across otherwise distinct sectors. The following subsections describe these major domains, outlining the protocols and mechanisms most exposed to quantum threats.

\subsubsection{Transport and Network Security}
\textbf{TLS~1.2/1.3.}
Transport Layer Security is the primary cryptographic protocol for securing web browsing, email submission and retrieval, directory access, and many IoT APIs. It provides server authentication, channel confidentiality, and integrity protection. Authentication relies on RSA/ECDSA X.509 certificates, while session keys are negotiated using Diffie–Hellman key exchange over finite fields (DHE) or elliptic curves (ECDHE)~\cite{rfc8446}. Under Shor’s algorithm, both digital signatures and ephemeral key exchanges are expected to become insecure: certificate forgery would enable server impersonation, and HNDL attacks would allow future decryption of captured sessions.

\textbf{QUIC.}
QUIC is a modern transport protocol designed for HTTP/3 and low-latency applications, integrating TLS~1.3 directly into its handshake \cite{rfc9000}. It inherits the same quantum exposures as TLS: HNDL attacks could affect ephemeral Diffie–Hellman exchanges and forgery would be possible against RSA/ECDSA certificates.

\textbf{IPsec/IKEv2.}
IPsec secures IP packets at the network layer, widely deployed in site-to-site VPNs and telecom backbones. IKEv2 negotiates session keys and authenticates peers using RSA/ECDSA certificates and (EC)DH \cite{rfc7296,ipsecme}. IKEv2 deployments using RSA/ECDSA and (EC)DH are exposed to signature forgery and HNDL under quantum-capable adversaries.

\textbf{SSH.}
The Secure Shell protocol is the standard for remote administration, file transfer, and tunneling in enterprise and cloud environments. It can authenticate servers and users using passwords, nevertheless, the best security practice is to use RSA, ECDSA, or Ed25519 public keys deriving session keys via X25519 key exchange \cite{rfc4253,openssh}.  In a post-quantum context, long-term RSA/ECDSA/Ed25519 keys could be forged using Shor’s algorithm, and captured sessions could be retrospectively decrypted (HNDL). In order to minimize the risk, modern OpenSSH versions \cite{openssh99} supports hybrid key exchange methods such as \texttt{sntrup761x25519-sha512@openssh.com} and \texttt{mlkem768x25519-sha256@openssh.com}, which combine classical X25519 with NIST-standardized post-quantum KEMs.

\textbf{WireGuard.}
WireGuard is a lightweight VPN protocol designed for performance and simplicity, relying on X25519 Diffie–Hellman for key exchange and AEAD for traffic protection \cite{wireguard}. It has no certificate-based component, so its exposure is purely HNDL.

\textbf{Wi-Fi Enterprise (EAP-TLS).}
Enterprise Wi-Fi networks use EAP-TLS for mutual authentication in 802.1X deployments. The protocol depends on X.509 RSA/ECDSA certificates and TLS key exchanges \cite{rfc9190}. With quantum capability, confidentiality of captured handshakes could be compromised (HNDL) and forged client or server certificates could undermine trust.

\textbf{WPA3-SAE.}
The WPA3 standard secures personal Wi-Fi using the Simultaneous Authentication of Equals password-authenticated key exchange, built on finite-field or elliptic-curve groups \cite{rfc7664} so, it is also vulnerable to HNDL.

\textbf{Bluetooth LE Secure Connections.}
BLE utilizes the Secure Connections (SC) model, relying on Elliptic Curve Diffie-Hellman (ECDH) with the P-256 curve for key establishment during the pairing phase \cite{bluetoothspec}. This process generates the Long Term Key (LTK) used for subsequent link-layer encryption and authentication. In the case of a quantum-enabled adversary the susceptibility of the P-256 primitive would expose BLE to two distinct risk vectors: recorded pairing handshakes would be subject to HNDL and the invalidation of the Diffie-Hellman exchange would enable active device impersonation and Man-in-the-Middle (MitM) attacks during the initial security association.

\textbf{Tor.}
The Tor anonymity network employs multi-hop routing across volunteer relays to provide resistance against direct source–destination correlation. Historically, its security model has relied on Ed25519 for relay identity and ntor (based on Curve25519) for circuit-extension handshakes \cite{tor-specs}. Tor deployments relying on Ed25519 and TLS are exposed to identity forgery and HNDL under future quantum-capable adversaries. Because Tor’s anonymity properties depend on the integrity of its cryptographic layers, compromise of the associated signature and key exchange primitives would reduce resistance to deanonymization, especially against adversaries capable of coordinated cryptographic and traffic-analysis attacks.
To mitigate these risks, the Tor Project is advancing PQ-Tor initiatives through the development of hybrid handshakes following Proposal 355 \cite{tor-prop355}, which layer ML-KEM alongside classical X25519. Furthermore, the migration to the Rust-based Arti implementation facilitates integration of post-quantum primitives. This aims to preserve the network's long-term integrity and anonymity against quantum-capable adversaries.

\textbf{Signal.}
Signal is a leading secure messenger, also forming the basis for WhatsApp’s cryptography. It historically used the X3DH and Double Ratchet protocols with X25519 and Ed25519 keys for session setup and authentication \cite{signal-specs}. These primitives would have been vulnerable to post-quantum attacks, enabling forgery of long-term identities and retrospective decryption of recorded sessions (HNDL). However, in 2023 Signal deployed the \emph{PQXDH} hybrid protocol, which combines the NIST-standardized ML-KEM with X25519 to provide quantum-resistant session establishment. This deployment represents one of the earliest large-scale integrations to integrate post-quantum protections in its core protocol \cite{signal-specs,signal-pqxdh}.

\subsubsection{Web and Mail Security}
\textbf{Web PKI (X.509).}
The global Web trust model binds domain names to public keys via CA-issued X.509 certificates \cite{rfc5280}, which use RSA or ECDSA for digital signatures. In a post-quantum setting, these signatures would be forgeable, enabling undetectable impersonation of HTTPS endpoints assuming a concurrent failure of the revocation infrastructure. In anticipation, multiple commercial CAs and PKI vendors have begun issuing pilot \emph{hybrid} or PQ-signed certificates under private trust or limited programs, pairing classical (RSA/ECDSA) and post-quantum signatures to preserve backward compatibility during transition.

\textbf{S/MIME (CMS).}
Secure/Multipurpose Internet Mail Extensions (S/MIME) provides end-to-end authentication and confidentiality for enterprise email using the Cryptographic Message Syntax (CMS, RFC~5652) as its container format \cite{rfc8551,rfc5652}. CMS defines \emph{SignedData} objects (digital signatures over message content and attributes) and \emph{EnvelopedData} objects (a content-encryption key, e.g., AES, wrapped for one or more recipients). In typical deployments, signatures use RSA or ECDSA, while confidentiality relies on either RSA key transport (the CEK encrypted to the recipient’s RSA public key) or ECC key agreement (ECDH per RFC~5753) to derive a key-encryption key (KEK) that wraps the CEK \cite{rfc5753}. Because S/MIME generally lacks forward secrecy (the CEK is bound to recipients’ long-term public keys), captured ciphertexts remain decryptable if those long-term keys are compromised in the future.

Quantum attacks would be expected to enable (i) \emph{forgery} of CMS \emph{SignedData} produced with RSA/ECDSA, and (ii) retrospective decryption (\emph{HNDL}) of CMS \emph{EnvelopedData} that used RSA key transport or ECDH for key establishment. The symmetric layer (e.g., AES-GCM) would remain viable but past messages protected under classical public-key mechanisms could be exposed.

\textbf{OpenPGP/GPG.}
OpenPGP is a foundational standard for securing email, software distribution, and data backups through asymmetric encryption and digital signatures \cite{rfc4880}. Historically reliant on RSA, ElGamal, and ECC primitives, OpenPGP implementations face a multi-faceted threat model in a post-quantum environment. These vulnerabilities manifest across three primary vectors: (i) the forgery of digital signatures, which undermines software supply chain integrity; (ii) the retrospective decryption of archived files and backups via HNDL; and (iii) the potential exposure of encrypted private key backups if they are protected by quantum-vulnerable algorithms.

\textbf{DKIM.}
DomainKeys Identified Mail (DKIM) is a standardized mechanism for authenticating the sending domain of email. When a message is transmitted, the sending mail transfer agent generates a digital signature over selected headers and the body, publishing the corresponding public key in the domain’s DNS zone. Receiving servers then retrieve the key via DNS and verify the signature, thereby linking the message to the purported sending domain rather than to an individual mailbox \cite{rfc6376}. Initially, DKIM deployments relied heavily on RSA-1024/2048 keys, with some support for ECDSA/Ed25519 signatures in modern profiles \cite{rfc8463}.

In a post-quantum setting, an adversary could derive the sender's private key from the publicly accessible DNS records. This would permit large-scale impersonation of trusted domains, bypassing spam and phishing filters that rely on DKIM validation, and undermining the integrity of SPF+DKIM+DMARC authentication chains. Because DKIM keys are persistent and globally queryable, they represent a static target for HNDL or offline key-recovery attacks.

\subsubsection{Identity and Single Sign-On}
\label{subsec:identity_sso}
\textbf{JWT/JWS (OAuth~2.0 and OIDC).}
JSON Web Tokens (JWT) are a compact JSON-based mechanism for conveying claims between parties, widely used for API authorization, mobile app sessions, and identity federation. They are typically signed with algorithms such as RS256 (RSA), PS256 (RSA-PSS), or ES256 (ECDSA), and may also be nested within OpenID Connect (OIDC) identity tokens \cite{rfc7515,rfc7518,oidf}.

The integrity and authenticity of JWTs depend on the security of the underlying asymmetric signature algorithms.
As JWTs function as bearer tokens, compromise of the associated signing algorithm under a cryptographically relevant quantum attack could allow an adversary to recover or forge the Identity Provider (IdP) signing key, which is typically exposed via a JSON Web Key Set (JWKS).

Such a compromise would permit the generation of forged tokens containing arbitrary claims or scopes, potentially enabling privilege escalation across services that trust the affected IdP.

Because many federated architectures rely on a limited set of IdP signing keys, these keys constitute centralized trust anchors whose compromise would have broad cross-service impact.

\textbf{SAML~2.0.}
The Security Assertion Markup Language (SAML) provides XML-based assertions used extensively in enterprise single sign-on (SSO) and cross-domain identity federation. Assertions convey authentication and attribute statements, while metadata exchanges establish trust between identity providers and service providers. These objects are secured with XML Digital Signatures (XMLDSIG) using RSA or ECDSA \cite{oasis-saml}. If those signatures could be forged by quantum-capable adversaries, they could inject or alter assertions, impersonating users across multiple domains.

The impact is especially severe in enterprise environments where SAML assertions are accepted by a wide range of critical applications, from HR systems to cloud productivity suites.

\textbf{WebAuthn/FIDO2.}
Web Authentication (WebAuthn) and the FIDO2 framework enable phishing-resistant multifactor authentication by binding credentials to specific hardware authenticators. Current implementations register ES256 (ECDSA over P-256) or EdDSA public keys and require signed challenges to complete authentication \cite{webauthn,w3c-fido2}. The ability to forge such authenticator responses by a quantum-enabled adversary would undermine the integrity of the challenge–response protocol. This would allow attackers to bypass MFA and take over accounts even without access to the user’s physical device. Because these standards are built on public-key signatures, the impact is significant given that WebAuthn is being adopted as a global passwordless standard.

\subsubsection{PKI and DNS Infrastructure}
\textbf{DNSSEC.}
The Domain Name System Security Extensions (DNSSEC) protect the integrity and authenticity of DNS responses by digitally signing zone records with algorithms such as RSA, ECDSA, or EdDSA \cite{rfc4033,rfc8080}. Validating resolvers verify these signatures against a chain of trust anchored at the root zone. In the presence of a quantum-capable adversary, the ability to forge these signatures would allow adversaries to conduct cache poisoning and zone takeover attacks that appear fully valid to resolvers. This would undermine the authenticity of hostnames and could redirect traffic for entire domains, affecting email delivery, HTTPS validation, and any service relying on DNS resolution.

\textbf{RPKI/BGPsec.}
The Resource Public Key Infrastructure (RPKI) underpins routing security by binding IP address prefixes to autonomous system numbers (ASNs) through signed Route Origin Authorizations (ROAs). BGPsec extends this model by requiring each AS in a path to sign its portion of the route advertisement \cite{rfc6480,rfc8205}. These mechanisms prevent prefix hijacking and route leaks when signatures are valid. The ability to forge ROAs or BGPsec path signatures for a quantum-capable adversaries would permit them to manipulate Internet routing tables without detection, potentially diverting traffic at global scale. Given the central role of BGP in the Internet backbone, such attacks could undermine routing integrity at scale under affected trust anchors.

\textbf{Certificate Transparency.}
Certificate Transparency (CT) provides publicly auditable, append-only logs of TLS certificate issuance, enabling detection of mis-issued or malicious certificates by browsers and monitoring services \cite{rfc9162}. Each log signs its entries and produces Signed Tree Heads (STHs) to prove consistency. In a post-quantum setting, forgery of log signatures would allow attackers to create fraudulent proofs of inclusion or conceal mis-issued certificates. This would erode the accountability of the Web PKI ecosystem, permitting undetected impersonation of domains even if browsers mandate CT compliance.

\subsubsection{Software Supply Chain}
\textbf{Code Signing.}
Code signing underpins trust in operating systems and applications by verifying that executables originate from a recognized vendor and have not been altered in transit. Mechanisms such as Microsoft Authenticode, Apple’s code-signing infrastructure, and Android APK signatures all rely on RSA or ECDSA certificates to validate software before execution \cite{ms-codesign,apple-codesign,apple-ecdsa,android-sig}. If such signatures could be forged, quantum-enabled adversaries could distribute malicious binaries that appear to be legitimate updates from trusted vendors. This would enable large-scale malware distribution through the very channels that users and administrators rely on for secure patching.

\textbf{Linux Package Managers.}
Linux distributions and open-source ecosystems rely heavily on package managers (e.g., APT, RPM, Flatpak, Snap) to deliver software and updates. These systems validate repository metadata using OpenPGP signatures \cite{debian-secure-apt, debian-repo-signing, redhat-rpm-signing}. Because package managers automatically resolve dependencies and push updates, the ability to forge maintainer or repository signatures by post-quantum adversaries would allow them to compromise thousands of systems simultaneously. The systemic nature of this risk makes package ecosystems particularly sensitive to signature integrity.

\textbf{Secure Boot.}
UEFI Secure Boot is designed to permit execution only of bootloaders, drivers, and kernels that are validated against firmware-resident signing keys, thereby establishing a chain of trust at system startup \cite{uefi-spec,uefi-secure-boot,uefi-secure-boot-chain}. The UEFI specification recommends use of RSA-2048 and allows ECDSA (e.g. ECDSA-384) for digital signatures on authorized images \cite{uefi-secure-boot-rsa-ecdsa,opencompute-secure-boot}. Under quantum-capable adversaries, forged bootloaders could bypass Secure Boot, installing stealthy pre-OS rootkits or persistent malware nearly invisible to higher-layer security tools. Because firmware and hardware root keys are difficult or impossible to replace in many systems, Secure Boot represents a primary firmware-level trust anchor in enterprise and consumer platforms.

\textbf{Sigstore/Cosign and Git.}
Modern DevSecOps pipelines extend signing practices beyond traditional binaries. Sigstore and its cosign tool provide signatures for container images, admission policies, and supply-chain attestations, while Git supports signed commits and tags using OpenPGP, X.509, or SSH keys \cite{sigstore}. These signatures enable continuous integration and delivery pipelines to enforce provenance and prevent tampering. In a post-quantum environment, forgery of these signatures would permit adversaries to inject malicious code into repositories, containers, or deployment pipelines while appearing to meet policy requirements. Such compromises could spread rapidly through automated build and deployment systems, undermining trust in the entire software supply chain.

\subsubsection{Documents and Records}
\textbf{PAdES, CAdES, XAdES.}
Electronic signature standards establish cryptographic mechanisms that enable legal recognition and integrity verification of digital documents across sectors such as business, government, and healthcare. PAdES (PDF Advanced Electronic Signatures) extends the PDF format with CMS-based signatures \cite{etsi_en319142_1}, CAdES (CMS Advanced Electronic Signatures) applies to general CMS structures \cite{etsi_en319122}, and XAdES (XML Advanced Electronic Signatures) secures XML documents \cite{etsi_en319132}. All rely on X.509 certificates and RSA/ECDSA (or equivalent public-key) signatures to ensure non-repudiation and long-term archival trust \cite{etsi_en319102_1}. In a post-quantum context, forged signatures could permit adversaries to tamper with contracts, invoices, or legal rulings while still appearing authentic \cite{etsi_tr103616}. Archival records, which may need to remain valid for decades, are especially at risk because retroactive validation depends on the enduring strength of the original signature algorithms.

\subsubsection{Payments and Identity Credentials}
\textbf{Contactless Payments.}
Contactless payment technologies, including EMV chip cards, NFC, and mobile wallets such as Apple Pay, rely on asymmetric cryptography to establish trust between the card and the terminal. Each transaction involves the generation of dynamic cryptograms, ensuring integrity and authenticity of the data exchanged \cite{emvcoAR2022,emvcoContactlessKernel2022}.

The EMV standard (Europay, MasterCard, and Visa) secures chip-based cards and point-of-sale transactions worldwide. While many transactions are processed online with issuer backend verification, EMV also specifies several \emph{offline authentication} mechanisms. These include Static Data Authentication (SDA), where the terminal verifies a signature over fixed card data; Dynamic Data Authentication (DDA), in which the card computes an RSA or ECC signature over a terminal challenge; and Combined DDA with Application Cryptogram (CDA), which signs both the challenge and the transaction details \cite{emvcoAR2024,emvcoECCFAQ2022}. In these modes, the card itself presents cryptographic evidence of authenticity to the terminal without needing to contact the issuer.

Historically, EMV deployed RSA with 1024- and 2048-bit keys for DDA and CDA, but since 2022 the EMV Contact Chip v4.4 and the EMV Contactless Kernel include support for Elliptic-Curve Cryptography (e.g., P-256) to improve efficiency and reduce key sizes \cite{emvcoECCFAQ2022,emvcoKL29_2025}. EMVCo issues periodic guidance on cryptographic key lengths to maintain security margins across the ecosystem.

Under quantum attack assumptions, an adversary could forge EMV signatures, creating counterfeit cards capable of passing offline authentication undetected. This risk is particularly acute in transport systems, retail, and regions with poor connectivity, where terminals routinely operate in offline mode \cite{emvcoPQC2020}. The long upgrade cycles of payment terminals and smart cards (often exceeding a decade) make the migration to post-quantum algorithms complex.

\textbf{ePassports/eIDs.}
Electronic passports and national eID cards are specified in the ICAO Doc~9303 standard, in particular Part~10 (Logical Data Structure, LDS) and Part~11 (Security Mechanisms) \cite{icao9303-10,icao9303-11}. Passive Authentication (PA) requires the terminal to verify RSA or ECDSA signatures made by the issuing state's Document Signer (DS) over the biometric and identity data groups stored in the chip. Each issuing country operates a Country Signing Certificate Authority (CSCA), which acts as the trust anchor for validating DS certificates during border inspection, and participating states publish their CSCA certificates through the ICAO Public Key Directory (PKD).

The initial secure channel between terminal and chip is established via Password Authenticated Connection Establishment (PACE), which replaced the legacy Basic Access Control (BAC) protocol and relies on Diffie-Hellman or Elliptic-Curve Diffie-Hellman (ECDH) key agreement \cite{icao9303-11}. PACE is directly exposed to quantum attacks: an adversary capable of breaking the underlying key agreement could decrypt intercepted border inspection sessions and expose biometric data in transit.

Modern ePassports implement Chip Authentication (CA) as the primary mechanism to prove chip genuineness and establish a chip-specific secure channel, superseding the older Active Authentication (AA) scheme. For documents containing sensitive biometric data such as fingerprints and iris scans, Terminal Authentication (TA) —part of the Extended Access Control (EAC) framework— additionally verifies that the reading terminal is authorized to access restricted data groups \cite{icao9303-11}. Both CA and TA rely on elliptic-curve cryptography and are therefore vulnerable to quantum attacks.

In a post-quantum context, a compromised or forged CSCA or DS certificate would enable adversaries to produce counterfeit ePassports or eIDs that pass automated border checks, undermining national identity infrastructures with broad cross-border implications including identity fraud and circumvention of entry restrictions. The HNDL threat is particularly relevant for the PACE channel: biometric data captured from intercepted inspection sessions today could be decrypted in the future.

\subsubsection{Blockchain and Cryptocurrencies}
\textbf{Bitcoin et al.}
Public blockchains such as Bitcoin and Ethereum rely on ECDSA \texttt{secp256k1} for transaction signatures \cite{secp256k1,ecdsaOriginal}. In these systems, control over funds is defined by the capability to generate valid signatures under the associated public key. In a post-quantum setting, quantum forgeries of ECDSA would enable attackers to generate signatures for any known public key, spending funds without access to the original private key. Beyond theft of cryptocurrency, this would also compromise smart contracts, decentralized finance applications, and blockchain-based identity systems built on the same signature primitives.

In Ethereum's case, in addition to ECDSA \texttt{secp256k1} governing externally owned accounts (EOAs), validator attestations in the Proof-of-Stake consensus layer rely on ellyptic curve BLS12-381 signatures \cite{eth2BLS} meaning that a quantum attack could simultaneously threaten both transaction integrity and consensus layer security.

Other widely adopted platforms face the same systemic risks. Ripple (XRP) supports multiple signature schemes, including \texttt{secp256k1}~\cite{secp256k1}, \texttt{secp256r1}~\cite{secp256r1}, and \texttt{Ed25519}~\cite{ed25519}, all of which are vulnerable under quantum-capable adversaries \cite{ripple2025}. Similarly, Solana (SOL) relies heavily on \texttt{Ed25519} signatures~\cite{ed25519} to achieve high-performance transaction validation, but this elliptic curve construction is equally vulnerable to quantum forgeries \cite{solana2025}. The widespread reliance on these elliptic-curve primitives across leading cryptocurrencies extends the scope of HNDL attacks, since adversaries can collect transaction data now and exploit it retroactively once quantum capabilities emerge.

It is important to note that the blockchain data structure itself—the append-only ledger secured by proof-of-work and SHA-256~\cite{sha256} (Bitcoin) or Keccak~\cite{keccak} (Ethereum)—is not expected to be fundamentally broken by quantum computers. While Grover’s algorithm reduces the effective security margin of hash functions, using SHA-256 or larger digests still provides sufficient protection. The primary exposure lies in the cryptographic layer that controls ownership and identity, not in the consensus mechanism that preserves ledger integrity. Because all historical blockchain data is immutable and publicly available, adversaries could exploit vulnerable addresses whose public keys are exposed with long-lived or reused public keys once quantum capabilities become practical.

\subsubsection{IoT and Industrial Systems}
\textbf{Consumer IoT.}
The consumer Internet of Things (IoT) ecosystem encompasses smart home platforms such as Matter, Apple HomeKit, and related device onboarding frameworks. These standards typically employ TLS channels secured by RSA/ECDSA X.509 device certificates for mutual authentication and encrypted telemetry exchange \cite{matter,apple-homekit}. In typical configurations, the mechanism restricts device enrollment to authorized endpoints and protects telemetry confidentiality. For a quantum-enabled attacker, the ability to forge device certificates would allow adversaries to enroll rogue devices into trusted ecosystems, potentially exfiltrating data or disrupting home automation workflows. Similarly, the confidentiality of stored or relayed telemetry traffic could be undermined by HNDL attacks.

\textbf{Industrial and Embedded Protocols.}
Operational Technology (OT) and Industrial Control Systems (ICS) rely on dedicated security standards to protect supervisory control and data acquisition (SCADA) environments. Protocols such as OPC Unified Architecture (OPC UA), IEC~62351/IEC~61850, and DNP3-SAv5 integrate TLS, X.509 certificates, and elliptic-curve Diffie–Hellman (ECDH) key exchanges to provide authentication and confidentiality for control commands, sensor telemetry, and automation processes \cite{iec62351,opcua-spec,ieee1815}. These mechanisms are critical in sectors such as power generation, manufacturing, transportation, and automotive systems, where unauthorized device access or falsified control signals can cause large-scale disruption or safety incidents.

\textbf{Automotive and V2X Security.}
Modern vehicles increasingly incorporate secure update frameworks, such as Uptane, and Vehicle-to-Everything (V2X) communication protocols. Uptane specifies signed metadata and cryptographic verification for over-the-air (OTA) software updates to Electronic Control Units (ECUs) \cite{uptane-spec}. V2X security is standardized in IEEE 1609.2 and ETSI ITS standards, both of which rely on ECDSA signatures over broadcast safety messages \cite{ieee1609-2, etsi-its}. A post-quantum compromise of these signatures would allow for the injection of fraudulent safety messages, potentially impacting safety-critical functions. Furthermore, the computational overhead of post-quantum signature verification presents a significant challenge for the low-latency requirements of real-time V2X communication.

The risk is also magnified by the extremely long deployment cycles of industrial and embedded platforms—often spanning 15–30 years—making rapid cryptographic migration difficult once vulnerabilities are exposed.

\medskip
A consolidated view of these technological domain-oriented risks, is provided in Table~\ref{tab:techdomains}.

%\section{Tabular Inventory of Quantum Risks}
% ======================= TABLE 1: TECHNOLOGY DOMAINS =======================
\begin{table*}[t]
\centering
\scriptsize
\caption{Inventory of Quantum Risks by Technology Domain}
\label{tab:techdomains}
\begin{tabularx}{\textwidth}{L{0.7 cm} L{5.0cm} L{2.2cm} L{3.9cm} X}
\toprule
\textbf{Domain} & \textbf{Technology} & \textbf{References} & \textbf{Classical Algorithm} & \textbf{Quantum Risk Type}\\
\midrule
\multirow[c]{9}{*}{\rotatebox{90}{\parbox{2.8cm}{\centering Transport\\Network}}}
  & TLS 1.2/1.3 (HTTPS, QUIC, SMTPS/IMAPS, LDAPs) & \cite{rfc8446,rfc9000} & RSA, ECDHE, ECDSA & HNDL (ECDHE); cert/signature forgery (RSA/ECDSA)\\
  & IPsec / IKEv2 & \cite{rfc7296} & RSA, (EC)DH, ECDSA & HNDL + authentication forgery\\
  & SSH & \cite{rfc4253,openssh99} & RSA, ECDSA, Ed25519, X25519 & HNDL (KEX); key/signature forgery\\
  & WireGuard & \cite{wireguard} & X25519 & HNDL (session decryption)\\
  & Wi-Fi Enterprise (EAP-TLS) & \cite{rfc9190} & RSA/ECDSA (X.509), ECDHE & HNDL; cert forgery\\
  & WPA3-SAE & \cite{rfc7664} & Finite-field/ECC PAKE & DL break $\to$ session compromise\\
  & Bluetooth LE Secure Connections & \cite{bluetoothspec} & ECDH P-256 & HNDL; impersonation\\
  & Tor (identities + TLS) & \cite{tor-specs} & Ed25519; RSA/ECDSA (TLS) & Identity forgery; HNDL\\
  & Signal (X3DH, Double Ratchet, PQXDH) & \cite{signal-specs,signal-pqxdh} & X25519, Ed25519; hybrid ML-KEM & Legacy HNDL + signature forgery; PQ-resistant for session keys (PQXDH)\\
\midrule
\multirow[c]{4}{*}{\rotatebox{90}{\parbox{1.0cm}{\centering Web\\Mail}}}
  & Web PKI (X.509) & \cite{rfc5280} & RSA, ECDSA & Certificate forgery\\
  & S/MIME (CMS) & \cite{rfc8551,rfc5652,rfc5753} & RSA, ECDSA, ECDH & SignedData forgery; HNDL decryption\\
  & OpenPGP / GPG & \cite{rfc4880} & RSA, ECC, ElGamal & Signature forgery; HNDL if EC-DH used\\
  & DKIM & \cite{rfc6376,rfc8463} & RSA; Ed25519/ECDSA & Domain signature forgery\\
\midrule
\multirow[c]{3}{*}{\rotatebox{90}{\parbox{1.0cm}{\centering Identity\\SSO}}}
  & JWT/JWS (OAuth2, OIDC) & \cite{rfc7515,rfc7518,oidf} & RS256/PS256/ES256 & Token forgery\\
  & SAML 2.0 (XMLDSIG) & \cite{oasis-saml} & RSA, ECDSA & Assertion forgery\\
  & WebAuthn / FIDO2 & \cite{webauthn,w3c-fido2} & ES256, EdDSA & Credential forgery\\
\midrule
\multirow[c]{3}{*}{\rotatebox{90}{Infra}}
  & DNSSEC & \cite{rfc4033,rfc8080} & RSA, ECDSA, EdDSA & Zone/RRset forgery\\
  & RPKI / BGPsec & \cite{rfc6480,rfc8205} & RSA, ECDSA & ROA/path forgery\\
  & Certificate Transparency & \cite{rfc9162} & RSA, ECDSA & Log/STH forgery\\
\midrule
\multirow[c]{5}{*}{\rotatebox{90}{\parbox{1.5cm}{\centering Software\\Supply Chain}}}
  & Code signing (OS, apps, APKs) & \cite{ms-codesign,apple-codesign,android-sig} & RSA, ECDSA (X.509) & Binary/firmware forgery\\
  & Package managers (APT, RPM, etc.) & \cite{debian-secure-apt,redhat-rpm-signing} & OpenPGP/GPG & Repository/package forgery\\
  & Secure Boot (UEFI) & \cite{uefi-spec} & RSA-2048; ECDSA-384 & Bootloader/firmware forgery\\
  & Sigstore / Cosign & \cite{sigstore} & X.509/OIDC (RSA/ECC) & Container/artifact forgery\\
  & Git (commits, tags) & \cite{git-docs} & OpenPGP; X.509; SSH & Commit/tag forgery\\
\midrule
\multirow[c]{4}{*}{\rotatebox{90}{\parbox{1.1cm}{\centering Docs\\Records}}}
  & PAdES & \cite{etsi_en319142_1,etsi_en319122,etsi_en319132,etsi_en319102_1} & RSA, ECDSA (CMS/X.509) & Document/signature forgery\\
  & CAdES & \cite{etsi_en319142_1,etsi_en319122,etsi_en319132,etsi_en319102_1} & RSA, ECDSA (CMS/X.509) & Document/signature forgery\\
  & XAdES & \cite{etsi_en319142_1,etsi_en319122,etsi_en319132,etsi_en319102_1} & RSA, ECDSA (CMS/X.509) & Document/signature forgery\\ \\
\midrule
\multirow[c]{4}{*}{\rotatebox{90}{\parbox{1.1cm}{\centering Payments \\ID creds.}}}
  & EMV chip / POS / NFC & \cite{emvcoAR2022,emvcoAR2024,emvcoECCFAQ2022,emvcoPQC2020} & RSA-2048, ECC P-256 & Offline Tx forgery; card impersonation\\
  & ePassports & \cite{icao9303-10,icao9303-11} & RSA, ECDSA & CSCA/DS forgery; ePassport cloning\\
  & eIDs & \cite{icao9303-10,icao9303-11} & RSA, ECDSA & CSCA/DS forgery; identity subversion\\ \\
\midrule
\multirow[c]{4}{*}{\rotatebox{90}{\parbox{1.1cm}{\centering Block- \\chain}}}
  & Bitcoin & \cite{ecdsaOriginal,secp256k1,ed25519,ripple2025} & ECDSA secp256k1 & Tx forgery; HNDL \\
  & Ethereum & \cite{ecdsaOriginal,secp256k1,ed25519} & ECDSA secp256k1 & Tx forgery; HNDL \\
  & Ripple (XRP) & \cite{ripple2025} & ECDSA secp256k1, secp256r1, Ed25519 & Tx forgery; HNDL \\
  & Solana & \cite{solana2025,ed25519} & Ed25519 & Tx forgery; HNDL \\
\midrule
\multirow[c]{2}{*}{\rotatebox{90}{\parbox{0.5cm}{\centering IoT}}}
  & Consumer IoT (Matter, HomeKit) & \cite{matter,apple-homekit} & TLS (RSA/ECDSA certs) & Cert forgery; HNDL session decryption\\
  & Industrial/Embedded (OPC UA, IEC 62351, DNP3, Uptane, V2X) & \cite{iec62351,opcua-spec,ieee1815,uptane-spec,ieee1609-2,etsi-its} & RSA/ECDSA; ECDH; Ed25519 (V2X) & Cert forgery; HNDL; malicious OTA/V2X injection\\
\bottomrule
\end{tabularx}
\end{table*}

% * * * * * * * * * * * *
\subsection{Operational Environment Risks}
This perspective examines how quantum threats manifest when cryptographic mechanisms are deployed within complete operational environments. Although the same primitives recur across multiple domains, their impact and criticality depend on how they are embedded into operating platforms, software stacks, and infrastructure layers. The following discussion outlines environment-specific patterns of exposure and the systemic risks that arise from these dependencies.

\textbf{Windows / Active Directory.}
Enterprise Windows infrastructures embed public-key cryptography into multiple critical control planes. Platform integrity relies on UEFI Secure Boot \cite{uefi-spec, uefi-secure-boot} and TPM-backed BitLocker \cite{ms-bitlocker, tcg-tpm}, where forged signatures on bootloaders or the Shor-based derivation of TPM-resident RSA keys would enable persistent, firmware-level compromise. Identity and directory services depend on AD CS \cite{ms-adcs} and Kerberos PKINIT \cite{rfc4556}, making certificate or ticket forgery capable of undermining domain-wide authentication. This extends the signature forgery risks identified in Section~\ref{subsec:identity_sso} to the Kerberos ticket-granting process. Remote access protocols (e.g., RDP \cite{ms-rdp}, SMB-over-QUIC \cite{ms-smb-quic}) and channel security (SChannel/TLS \cite{rfc8446}) rely on RSA/ECDSA, exposing them to both HNDL and active impersonation risks. Furthermore, smart card logon (PIV \cite{fips201-3}) and driver/code-signing chains \cite{ms-codesign} extend the exposure to physical credentials and kernel-level system updates. The result is a highly interdependent trust model where the compromise of a single Certificate Authority (CA) or hardware anchor could compromise multiple security functions within the domain’s trust boundary.

\textit{Linux and Open-Source Ecosystems.}
Open-source ecosystems distribute trust across decentralized projects, yet nearly all foundational security mechanisms rely on quantum-vulnerable asymmetric primitives. Platform integrity is anchored in UEFI Secure Boot \cite{uefi-secure-boot} and kernel module signing \cite{linux-kernel-modsign}, while secure communications and cluster orchestration frameworks depend on RSA/ECDSA certificates and DH-based key exchanges \cite{rfc4253, rfc7296, k8s-docs}. The software supply chain is particularly critical; package managers, container registries, and developer tooling rely on maintainer signatures that, if forged, could enable ecosystem-wide poisoning \cite{debian-secure-apt, sigstore}.

While projects like Open Quantum Safe have initiated PQC integration \cite{oqs}, the heterogeneity and decentralization of the Linux and OSS landscape present significant challenges for systemic, consistent migration.

\textbf{macOS / Apple Ecosystem.}
Apple’s desktop ecosystem relies on a vertically integrated trust model where both hardware and software roots of trust depend on classical cryptography. Secure Boot, System Integrity Protection (SIP) \cite{apple-sip}, and the Sealed System Volume (SSV) \cite{apple-sealed} use RSA/ECDSA-based signatures to validate boot stages and prevent unauthorized modifications. App notarization and kext signing \cite{apple-code-sign, apple-codesign} rely on Apple’s own code-signing infrastructure, exposing the execution environment to forged signature risks. Identity services (iCloud Keychain, Apple ID logins) and native VPN/TLS stacks (Secure Transport, NetworkExtension) inherit classical crypto dependencies that are vulnerable to HNDL attacks \cite{rfc8446}. As with iOS, the Secure Enclave provides hardware-backed key storage but issues classical key material \cite{apple-platform-security}. The centralized, curated nature of macOS security helps facilitate fast updates—but any compromise of Apple's signing infrastructure could affect a large number of endpoints at once.

\textbf{BSD / Unix Derivatives.}
BSD-based systems (FreeBSD, OpenBSD, and HardenedBSD) prioritize code correctness and minimalism, yet they also depend heavily on classical cryptographic primitives. Secure boot mechanisms (UEFI-based \cite{freebsd-uefi-docs} or loader.conf-initiated) are limited in some distributions, but when present, depend on RSA/ECDSA. SSH is the dominant remote access method \cite{rfc4253}, relying on classical key exchanges and signature verification. FreeBSD’s pkg manager \cite{freebsd-pkg} uses RSA/ECDSA signatures to verify manifests and binaries, meaning forged keys could compromise system integrity. OpenBSD’s signify tool \cite{openbsd-signify} and source code auditing practices \cite{openbsd-audit} help mitigate some risks, but the fundamental reliance on classical signatures leaves software distribution vulnerable to HNDL and forgery attacks. PQ migration is in early stages within select BSD communities (e.g., testing of libOQS \cite{oqs}), but lack of vendor support and decentralized development hinder rapid adoption. As BSD is widely used in embedded systems, firewalls, and routers \cite{pfSense, hardenedbsd-docs}, these quantum-exploitable weaknesses could persist for years.

\textbf{Mobile Platforms (Android/iOS).}
Mobile platforms concentrate security within hardware-backed trust anchors that protect billions of users. Verified boot chains, such as Android AVB 2.0 \cite{android-avb} and Apple's iBoot \cite{apple-platform-security}, alongside application distribution frameworks like APK signatures \cite{android-sig} and App Store signing \cite{apple-codesign}, depend heavily on RSA/ECDSA primitives. Consequently, post-quantum forgeries could bypass both operating system integrity and application vetting processes.

Hardware keystores, including StrongBox \cite{android-keystore} and the Secure Enclave \cite{apple-platform-security}, currently issue classical key material, leaving long-term identity credentials at risk. Mobile TLS stacks (e.g., Conscrypt, BoringSSL, and Secure Transport \cite{rfc8446}) inherit HNDL exposure from their reliance on classical handshakes. Furthermore, push notification services (e.g., FCM \cite{fcm}, APNs \cite{apns}) and payment infrastructures (Apple Pay \cite{apple-pay}, Google Pay \cite{google-pay}) utilize JWTs or ECC-based tokens \cite{rfc7515, rfc7518}, which are susceptible to forgery, potentially enabling credential theft or fraudulent financial transactions. While messaging protocols like Signal PQXDH \cite{signal-pqxdh} and iMessage PQ3 \cite{apple-pq3} have pioneered PQ/hybrid deployments, platform-wide migration across the OS and third-party ecosystems remains incomplete.

\textbf{Cellular Networks.}
Telecommunications infrastructures exhibit a heterogeneous security posture, characterized by the relative quantum-resilience of symmetric-key subscriber authentication and the critical vulnerability of asymmetric control-plane signaling. Subscriber authentication protocols, including AKA, EPS-AKA, and 5G-AKA \cite{ts33401, ts33501}, rely primarily on symmetric ciphers, which remain secure with sufficiently long keys in a post-quantum environment. However, inter-operator trust and control-plane signaling rely heavily on asymmetric primitives.

NDS/IP tunnels secured by IKEv2 \cite{rfc7296, ts33210}, SUCI identity concealment via ECIES, and TLS-protected service-based architectures \cite{rfc8446} are all vulnerable to HNDL or certificate forgery. Additionally, JWT/OAuth tokens used within network functions \cite{rfc7515} represent a significant forgery vector. Under a cryptographically relevant quantum adversary model, forged operator or equipment credentials could enable impersonation of legitimate infrastructure elements like rogue base stations, enabling traffic interception and degrading trust across roaming and interconnect domains. Given that telecom standards operate on multi-year release cycles and equipment lifecycles span decades, the window of systemic exposure will be operationally challenging to mitigate\cite{gsma-pqc}.

\textbf{Industrial / Operational Technology (OT).}
Industrial control and critical infrastructure (CI) environments exhibit pervasive dependencies on quantum-vulnerable asymmetric primitives. PKI hierarchies bind vendor, plant, and device identities via RSA/ECDSA, exposing plants to device impersonation if certificates can be forged. Core SCADA and OT protocols—OPC UA \cite{opcua-spec}, IEC~62351/61850 \cite{iec62351,ieee-61850}, DNP3-SAv5 \cite{dnp3-sav5}, CIP \cite{odva-cip}, PROFINET \cite{profinet-security}, DLMS/COSEM \cite{dlms-cosem}—rely on TLS and X.509, introducing HNDL and forgery risks into environments where uptime and safety are paramount. Brokered telemetry systems (MQTT \cite{oasis-mqtt}, AMQP \cite{oasis-amqp}) face similar vulnerabilities when TLS credentials are broken. Edge gateways—responsible for secure boot, firmware signing, and updates \cite{uefi-secure-boot}—extend exposure to the device lifecycle. Given the decadal operational lifecycles of OT hardware, these quantum-vulnerable dependencies represent a persistent, long-term risk to systemic infrastructure integrity \cite{ics-pqc}.

\textbf{Automotive Systems.}
Although the preceding discussion examined automotive security mechanisms at the protocol and standards level, their practical impact depends on how these cryptographic trust anchors are embedded within complete vehicular systems. Modern vehicles rely on asymmetric cryptography to enforce software integrity, identity, and update control. Over-the-Air (OTA) frameworks such as Uptane \cite{uptane-spec} validate signed firmware using RSA or ECDSA. In-vehicle networks (e.g., CAN, CAN-FD, Automotive Ethernet) lack intrinsic authentication and therefore depend on gateway-enforced trust mechanisms. Externally, V2X standards such as IEEE~1609.2 \cite{ieee1609-2} and ETSI ITS \cite{etsi-its} use ECC-based certificates for vehicular identity and message integrity, exposing them to signature forgery under quantum-capable adversaries. Given the long lifecycle of vehicles and limited upgrade agility, these dependencies introduce prolonged exposure if the underlying asymmetric primitives are weakened.

\medskip
The environment-oriented analysis shows that quantum risks vary significantly with system context. In enterprise infrastructures, compromise of directory services or PKIs could undermine entire domains; in open-source ecosystems, package managers and software supply chains form systemic weak points; in mobile platforms, verified boot and application signing constitute single points of failure; in telecommunications, inter-operator trust and subscriber privacy rely on vulnerable public-key mechanisms; in industrial and automotive environments, device lifecycles and safety-critical processes magnify the consequences of quantum compromise. These findings emphasize that quantum exposure is shaped not only by the cryptographic primitives themselves, but also by their integration into operational environments.

A consolidated overview of these environment-specific risks is provided in Tables~\ref{tab:op-environments}.

% ======================= TABLE II: Operational Environments extended references =======================
\renewcommand{\thetable}{II}
\begin{table*}[!t]
\centering
\begingroup
\scriptsize
\setlength{\tabcolsep}{3.5pt}
\renewcommand{\arraystretch}{1.03}
\caption{Inventory of Quantum Risks by Operational Environment}
\label{tab:op-environments}
\begin{tabularx}{\textwidth}{L{1.8cm} L{2.5cm} L{2.0cm} L{2.0cm} L{2.8cm} X L{2.0cm}}
\toprule
\textbf{Environment} & \textbf{Crypto Uses} & \textbf{Vulnerable Algorithms} & \textbf{Quantum Risk Types} & \textbf{Compromise Effects} & \textbf{Example Technologies} & \textbf{References} \\
\midrule

\textbf{Windows / AD} & Boot integrity, directory services, remote access, code signing & RSA, ECDSA, DH, PKINIT & HNDL, signature forgery, impersonation & Domain-wide compromise, persistent malware, driver injection & UEFI Secure Boot, BitLocker, AD CS, Kerberos PKINIT, RDP, SMB, SChannel, Smartcard Logon & \cite{uefi-secure-boot,ms-bitlocker,ms-adcs,rfc4556,ms-rdp,ms-smb,ms-smb-quic,rfc8446} \\
\midrule

\textbf{Linux (OS layer)} & Boot chain, kernel integrity, remote login, package and repo signing & RSA, ECDSA, X25519, DH & HNDL, signature forgery, supply chain injection & Malicious kernel modules, repo poisoning, developer impersonation & UEFI Secure Boot, Linux kernel module signing, OpenSSH, APT, RPM, TPM subsystem, Git, Sigstore, OCI registries & \cite{uefi-secure-boot,linux-kernel-modsign,rfc4253,openssh,debian-secure-apt,redhat-rpm-signing,tpm-linux-docs,git-docs,sigstore,oci} \\
\midrule
\textbf{macOS / Apple} & Secure boot, app notarization, OS integrity, ID services & RSA, ECDSA & HNDL, signature forgery & Forged apps/kexts, bypass of SIP/SSV, stolen Apple ID creds & SIP, Sealed System Volume, App Notarization, Secure Enclave, iCloud Keychain & \cite{apple-sip,apple-sealed,apple-code-sign,apple-codesign,apple-platform-security,rfc8446} \\
\midrule

\textbf{BSD / Unix} & Boot loaders, package signing, SSH remote access & RSA, ECDSA & HNDL, forgery & Compromised packages, router/firewall impersonation & FreeBSD pkg, OpenBSD signify, loader.conf boot, pfSense, HardenedBSD & \cite{freebsd-pkg,openbsd-signify,freebsd-uefi-docs,rfc4253,pfSense,hardenedbsd-docs,openbsd-audit} \\
\midrule

\textbf{Mobile Platforms} & Verified boot, app distribution, keystore, payments & RSA, ECDSA, ECC tokens & HNDL, signature forgery, token forgery & Malicious apps, bypass of verified boot, fraudulent payments & Android AVB~2.0, APK Signature Scheme, iBoot, App Store signing, StrongBox, Secure Enclave, Apple Pay, Google Pay, PQ3 (iMessage), PQXDH (Signal) & \cite{android-avb,android-sig,apple-codesign,apple-platform-security,android-keystore,apple-pay,google-pay,fcm,apns,rfc7515,rfc7518,apple-pq3,signal-pqxdh} \\
\midrule

\textbf{Cellular Networks} & Subscriber auth, inter-operator tunnels, service-based arch. & RSA, ECDSA, ECIES & HNDL, certificate forgery, impersonation & Fake subscribers, rogue base stations, MITM between operators, SBA disruption & EPS-AKA, 5G-AKA, IKEv2, NDS/IP, SUCI, SBA TLS, OAuth/JWT tokens & \cite{ts33401,ts33501,ts33210,rfc7296,rfc8446,rfc7515,gsma-pqc} \\
\midrule

\textbf{Industrial / OT} & PKI for device identity, SCADA comms, telemetry security & RSA, ECDSA, DH, X.509 & HNDL, certificate forgery & Device spoofing, falsified control signals, unsafe shutdowns & OPC UA, IEC~62351/61850, DNP3-SAv5, CIP, PROFINET, DLMS/COSEM, MQTT, AMQP, Edge gateways & \cite{opcua-spec,opcua-pki,iec62351,ieee-61850,ieee1815,dnp3-sav5,odva-cip,profinet-security,dlms-cosem,oasis-mqtt,oasis-amqp} \\
\midrule

\textbf{Automotive} & OTA updates, in-vehicle communications, ECU firmware validation, V2X messaging & RSA, ECDSA & HNDL, forgery & Malicious firmware, fake V2X safety alerts, ECU compromise & Uptane, CAN-FD, CAN gateways, IEEE~1609.2, ETSI ITS security, AUTOSAR & \cite{uptane-spec,ieee1609-2,etsi-its} \\

\bottomrule
\end{tabularx}
\endgroup
\end{table*}

% CEB HASTA ACA!!!!
%-----------------------------------------------------------------------
\section{Conclusion}
\label{sec:conclusion}
This paper has presented a systematic inventory of cryptographic vulnerabilities across modern digital infrastructures, categorized by two primary risk vectors: Harvest Now, Decrypt Later (HNDL) and signature forgery. Through a dual-layered taxonomy of technology domains and operational environments, this analysis documents the reliance of current protocols on asymmetric primitives susceptible to Shor’s algorithm.

Tables~\ref{tab:techdomains} and~\ref{tab:op-environments} consolidate these findings, mapping the quantum threat across a spectrum that ranges from Internet-scale transport protocols to constrained embedded systems. The inventory indicates that exposure to quantum-vulnerable primitives is present across consumer, enterprise, and critical infrastructure sectors. It is noteworthy to discover that there are a substantial number of protocols and systems that rely on public-key cryptography, which multiplies and extends vulnerability in the case of adversaries with quantum capabilities.

\section{Future work}
\label{sec:future-work}
Future work may include:
\begin{enumerate}
\item \textbf{Graph-Based Dependency and Impact Modeling:} Application of graph theory to model the hierarchical trust chains identified in this inventory. Future research could quantify the propagation of compromise resulting from specific primitive failures—identifying critical architectural nodes, such as Identity Providers or Root Certificate Authorities, where a single asymmetric vulnerability facilitates a broad-scale compromise of the security architecture.

\item \textbf{Extension to Specialized Operational Contexts:} Expanding the inventory to specialized sectors with distinct architectural constraints, such as satellite communications, subsea infrastructure, and implantable medical devices. These environments often utilize non-standard protocol implementations and legacy primitives that may exhibit unique vulnerability profiles not captured in general-purpose computing frameworks.
\end{enumerate}

%------------------------------------------------------------------------------
\section*{Acknowledgment}
The author acknowledge the use of generative AI (OpenAI) for discovery assistance and linguistic polishing. The technical taxonomy, the mapping of cryptographic dependencies, and the resulting engineering inventory were developed by the author, who takes full responsibility for the integrity and accuracy of the final work.

The author would also like to thank Dr.~Nina Bindel and Dr.~Fernando Virdia for their valuable reviews of this work. Their constructive feedback and insightful suggestions regarding post-quantum security parameters greatly improved the clarity, accuracy, and overall quality of the paper.

\bibliographystyle{IEEEtran}
\bibliography{mapping_quantum_threats}

\end{document}